\documentclass[]{aastex631}

\usepackage{amsmath}
\usepackage{amssymb}
\usepackage{graphicx}
\usepackage{multirow}
\usepackage{natbib}

\usepackage[english]{babel}
\usepackage[utf8]{inputenc}
\usepackage{algpseudocodex}

\newcommand{\arc}{\stackrel\frown}

\newcommand{\red}[1]{{#1}}

\shortauthors{C.-C. He}

\begin{document}

\title[Analytic TPCFs]{A Fast and Accurate Analytic Method of Calculating Galaxy Two-point Correlation Functions}

\correspondingauthor{Chong-Chong He}
\email{che1234@umd.edu}

\author[0000-0002-2332-8178]{Chong-Chong He}
\affiliation{Department of Astronomy, University of Maryland, College Park, MD 20742-2421}

\begin{abstract}
  We have developed a new analytic method to calculate the galaxy two-point correlation functions (TPCFs) \red{accurately} and efficiently, applicable to surveys with finite, regular, and mask-free geometries.
  We have derived simple, accurate formulas of the normalized random-random pair counts $RR$ as functions of the survey area dimensions. We have also suggested algorithms to compute the normalized data-random pair counts $DR$ analytically.
  With all edge corrections fully accounted for analytically, our method computes $RR$ and $DR$ with perfect accuracy and zero variance in $O(1)$ and $O(N_{\rm g})$ time, respectively.
  \red{We test our method on a galaxy catalogue from the EAGLE simulation. Our method calculates $RR+DR$ at a speed 3 to 6 orders of magnitude faster than the brute-force Monte Carlo method and 2.5 orders of magnitude faster than tree-based algorithms.
  For a galaxy catalogue with 10 million data points in a cube, this reduces the computation time to under 1 minute on a laptop.}
  Our analytic method is favored over the traditional Monte Carlo method whenever applicable.
  \red{Some applications in the study of correlation functions and power spectra in cosmological simulations and galaxy surveys are discussed. However, we recognize that its applicability is very limited for realistic surveys with masks, irregular shapes, and/or weighted patterns.}
\end{abstract}

\section{Introduction}

The two-point correlation function (TPCF) $\xi(r)$ has been the primary tool for quantifying large-scale cosmic structures \citep{Peebles1980}.
$\xi(r)$ is defined as the fractional increase relative to a random Poisson distribution in the probability \(\delta P\) of finding objects in two volume elements \(\delta V_1\) and \(\delta V_2\) separated by distance \(r\):
\begin{equation}\label{dP}
\delta P = [1 + \xi(r)] n^2 dV_1 dV_2,
\end{equation}
where \(n\) is the mean number density of objects (galaxies or dark matter halos).

The Fourier transform of $\xi(r)$ is the galaxy power spectrum, which is often used to describe the structure of the Universe \citep{Peebles1980}.
Starting from \cite{Eisenstein2005}, $\xi(r)$ has become a popular tool for the detection of the galaxy clustering signal at 150 Mpc known as the baryon acoustic oscillations. It is a signature of the density difference that arose from the first million years of the Universe.

Given a survey or simulation containing the 3D coordinates of all galaxies,
the most straightforward way to estimate $\xi(r)$ is to take the ratio of the number of data-data pairs, $DD$, to that expected from a random distribution in the same area, the random-random pair counts $RR$, properly normalized, minus one:
\begin{equation}\label{eq2}
  \xi_{0}(r) = \frac{\arc{DD}}{\arc{RR}} - 1.
\end{equation}
This is known as the natural estimator.
Other estimators involving cross-pair separation count $DR$ between the data set and random set have been proposed to reduce the estimation variance, notably induced by edge effects.
The Landy-Szalay estimator \citep{Landy1993},
\begin{equation}\label{eq:LS}
  \xi_{\rm LS}(r) = \frac{\arc{DD} - 2\arc{DR} + \arc{RR}}{\arc{RR}},
\end{equation}
is the most commonly used estimator because it has minimal variance and converges to the direct estimate the fastest \citep[][]{Kerscher2000}.

For surveys with non-periodical boundary conditions, it is a common practice is measure the average available bin using Monte Carlo integration, by generating a comparison random distribution of a large number of points over the same survey area.
To reduce statistical fluctuations, it is standard to use densely populated random fields, usually 50 times denser than the survey population. Because the number of computations needed to measure the separations between $N$ objects scales as $N^2$, the random-random pair counts $RR$ dominates the computing time.
For a large survey or simulation, especially, the computation of $RR$ can be extremely time-consuming.

\red{
Tree-based algorithms and code exist that are much faster than the brute-force method \citep{Moore2001, Jarvis2004, Zhang2005}. {\sc TreeCorr}, for instance, is a widely used tree-based code that computes TPCFs in $O(N \log N)$ time \citep{Jarvis2004}. This improvement in speed, however, comes with a sacrifice in accuracy. As pointed out in \cite{Siewert2020}, {\sc TreeCorr} shows noticeable errors in the computation of angular correlation function under any setting that has a significant advantage in speed.
Computationally efficient approaches to calculating TPCFs have also been proposed \citep{Demina2018,Keihanen2019}. However, they reply on a populated random catalogue and their efficiency and accuracy are limited.
}

\red{
The random-random pair counts $\arc{RR}(r)$ is a purely geometrical quantity that depends only on the geometry of the galaxy catalogue and the radial selection function in the case of cosmological survey. The data-random pair counts $\arc{DR}(r)$ is also a well-defined quantity given a galaxy population in a given geometry.
However, analytic methods of estimating pair counts that apply to finite geometries are very sparse in the literature, with only two such works found to our best knowledge. \cite{Demina2018} developed a semi-analytic method to compute $RR$ and $DR$ in two of the three dimensions, but still using a random catalogue to account for angular correlations.
\cite{Breton2021} proposed a method to estimate the angular pair counts based on analytic integral expressions. This scheme, however, relies on conducting numerical integrations of terms including the angular selection function over the full sky, which is non-trivial for surveys in subregions of the sky or surveys with masks.
}

In this paper, we derive analytic formulas of $\arc{RR}(r)$ with simple closed-form expressions and analytic algorithms of $\arc{DR}(r)$ with perfect accuracy and zero variance, applicable to \red{mask-free} surveys with regular geometries.
\footnote{A code written in Python is published (doi:10.5281/zenodo.5201479), as developed on \url{https://github.com/chongchonghe/analytic-2pcf/tree/v1.0.0}.}
\red{We apply these formulas and algorithms to a mock galaxy catalogue from the EAGLE simulation \citep{McAlpine2016} and assess the accuracy and speed of the method compared to traditional methods that utilize random catalogues.}

The remainder of this article is organized as follows. In Section~\ref{sec:1} we derive analytic formulas of $\arc{RR}(r)$ for four groups of survey geometries. In Section~\ref{sec:dis} we estimate the fractional corrections on $\arc{RR}(r)$ caused by edge corrections. In Section~\ref{sec:dr} we present algorithms to compute $\arc{DR}(r)$ analytically. We compare our analytic method with the traditional Monte Carlo method in Section~\ref{sec:3}, followed by summaries and discussions in Section~\ref{sec:sum}.

\section{Analytic Formulas of $\arc{RR}$}
\label{sec:1}

We start with deriving formulas of the random-random pair counts $\arc{RR}(r)$ of a mask-free survey with boundaries, taking into account all edge corrections. The survey geometries we consider here are rectangles, cuboids, circles, and spheres.
For each case, we first compute the bulk random-random pair counts ignoring any edge effects.
Then we do edge corrections, excluding the pairs where the second data point is outside of the survey region. This method of calculating $\arc{RR}$ is equivalent to the Monte Carlo method with infinite number of random points. With all done analytically, this algorithm of computing $\arc{RR}(r)$ has $O(1)$ scaling, i.e., it does not scale with the number of data points at all.

It is necessary to point out that the number of random-random pair counts from $r$ to $r + dr$ is $\arc{RR}(r) dr$. One would need to integrate $\arc{RR}(r)$ over $r$ from $r_1$ to $r_2$ in order to get a bined $\arc{RR}(r)$.

\subsection{Rectangles}

\begin{figure}
  \centering
  \includegraphics[width=\linewidth]{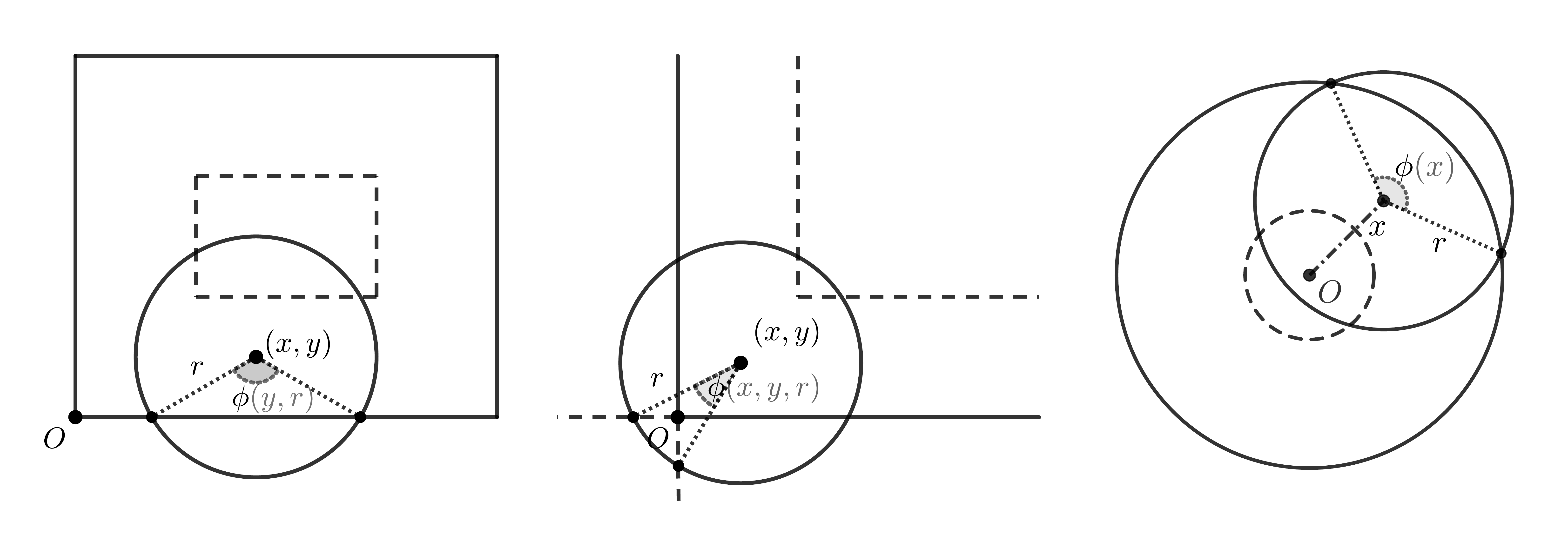}
  \caption{\label{fig:1.1} Diagrams showing edge corrections in the calculation of random-random and data-random pair counts. From left to right are corrections of the edges of a rectangle, the corners of a rectangle, and the edges of a circle. Cuboids and spheres are similar.
  }
\end{figure}

In the first geometry, we consider a rectangular region with sides $a$ and $b$. We compute $\arc{RR}(r)$ in three steps where we take care of 1) the whole region ignoring boundaries, 2) corrections of the edges, and 3) corrections of the corners.

In step I, we consider all possible pairs where the first object is drawn inside the rectangle. The number of such pairs with separations between $r$ and $r+dr$ is simply
\begin{equation}\label{eq:1.1}
  \Theta_{\rm I}(r) \ dr = (a b) (2 \pi r) n^2 \ dr = 2\pi abr n^2 \ dr ,
\end{equation}
where $n$ is the number density of objects. For simplicity, the $dr$ is omitted on both sides of the formula for the rest of this article.

In step II, we exclude all the pairs where the second point is in either $x<0$, or $x>a$, or $y<0$, or $y>b$.  The number of such pairs for $y<0$ is
\begin{equation}\label{eq:1.2}
  \Theta_{\rm II, y<0}(r) = \int_{0}^r dy \ \phi(y) a r \red{n^2} = 2 a r^2 n^2,
\end{equation}
where
\begin{equation}\label{eq:1.2.2}
  \phi(y) = 2 \cos^{-1}(\frac{y}{r})
\end{equation}
is the angle of the arc at $y<0$ (Fig.~\ref{fig:1.1}, left).
Similarly,
\begin{align}\label{}
  \Theta_{\rm II, y>b}(r) &= 2 a r^2 n^2,\\
  \Theta_{\rm II, x<0}(r) &= 2 b r^2 n^2,\\
  \Theta_{\rm II, x>a}(r) &= 2 b r^2 n^2.
\end{align}
The total number of pairs to exclude in this step is therefore
\begin{equation}\label{}
  \Theta_{\rm II}(r) = \Theta_{\rm II, x<0}(r) + \Theta_{\rm II, x>a}(r) + \Theta_{\rm II, y<0}(r) + \Theta_{\rm II, y>b}(r) = 4(a + b) r^2 n^2.
\end{equation}

In step III, we add back the pairs that are excluded twice. These are the pairs where the second object is either $x<0$ \& $y<0$, or $x>a$ \& $y<0$, or $x<0$ \& $y>b$, or $x>a$ \& $y>b$. The number of pairs in these four cases is all identical and their total is given by
\begin{equation}\label{eq:1.4}
  \Theta_{\rm III}(r) = 4 \ \int_{0}^r dx \int_0^{\sqrt{r^2 - x^2}} dy \ \phi(x,y) r = 2 r^3 n^2,
\end{equation}
where
\begin{equation}\label{eq:1.5}
  \phi(x,y) = \frac{\pi}{2} - \sin^{-1} \frac{x}{r} - \sin^{-1} \frac{y}{r},
\end{equation}
is the angle of the arc at $x<0 \ \& \ y<0$ (Fig.~\ref{fig:1.1}, center).

Finally, combining steps I, II, and III, the total number of valid random-random pairs for the rectangular field with sides $a$ and $b$ and object density $n$ is given by
\begin{align}
  \Theta(r) &= \Theta_{\rm I} - \Theta_{\rm II} + \Theta_{\rm III} \\
  &= \left[ 2\pi abr - 4(a+b)r^{2} + 2 r^3 \right] n^2,
  \label{eq:1.8}
\end{align}
The normalized random-random pair count is therefore
\begin{equation}\label{eq:1.9}
  \arc{RR}_{\rm rect}(r) = \frac{\Theta(r)}{(abn)^{2}} = \frac{2\pi}{ab} r - \frac{4(a+b)}{a^2 b^2}r^{2} + \frac{2}{a^{2} b^2} r^3,
\end{equation}
for $\red{r \le a \le b}$.
\red{
The antiderivative is given below for an easy calculation of its integration over $r$.
\begin{equation}\label{eq:1.9.1}
  {\cal F}[\arc{RR}_{\rm rect}(r)] = \frac{\pi}{ab} r^2 - \frac{4(a+b)}{3a^{2} b^2} r^3 + \frac{1}{2 a^{2} b^2} r^4.
\end{equation}}
In a special case where $a = b$, Eq.~(\ref{eq:1.9}) becomes
\begin{equation}\label{eq:1.10}
  \arc{RR}_{\rm square}(r) = \frac{2\pi}{a^2} r - \frac{8}{a^3}r^{2} + \frac{2}{a^{4}} r^3.
\end{equation}

\red{We further discuss the situation when $a < r \le b$. In this case, $\Theta_{\rm I}$ is unchanged. In the calculation of $\Theta_{\rm II}$, integrations over $y$ is unchanged and yield $4 a r^2 n^2$. In the integration over $x$, however, the upper limit of the integral in Eq.~(\ref{eq:1.2}) needs to be replaced by $a$. Summing up, it yields
\begin{equation}\label{}
  \Theta_{\rm II}(r) = \left[ 4 a r^2 + 4 b r \left(r - \sqrt{r^2 - a^2} + a \cos^{-1}\frac{a}{r} \right) \right] n^2.
\end{equation}
In $\Theta_{\rm III}$, similarly, the upper limit in the first integral needs to be changed to $a$ in the integration over $x$, yielding
\begin{equation}\label{}
  \Theta_{\rm III}(r) = \left[ 1 + (2 - \frac{a}{r})\frac{a}{r} \right] r^3 n^2.
\end{equation}
Adding together and conducting normalization, we find
\begin{equation}\label{eq.1.9v2}
  \arc{RR}_{\rm rect}(r) = \frac{2\pi}{ab} r - \frac{4 a + 4 b \left(1 - \sqrt{1 - (\frac{a}{r})^2} + \frac{a}{r} \cos^{-1}\frac{a}{r} \right)}{a^2b^2} r^2 + \frac{1 + \left( 2 - \frac{a}{r} \right) \frac{a}{r}}{a^2b^2} r^3
\end{equation}
for $a < r \le b$.}

\subsection{Cuboids}

Now we consider a cuboid in 3D space with sides $a$, $b$, and $c$. We calculate $\arc{RR}(r)$ in four steps where we take care of 1) the whole region ignoring edge effects, 2) corrections from the faces, 3) corrections from the edges, and 4) corrections from the corners.

In the step I, consider all possible pairs where the first point is inside, ignoring any edge effects,
\begin{equation}\label{}
  \Theta_{\rm I}(r) = (abc) (4\pi r^{2}) n^2 = 4 \pi abc r^2 n^2.
\end{equation}

In step II, we exclude the pairs where the second point is outside one of the faces of the cuboid. There are a total of 6 cases, corresponding to the 6 faces of a cuboid. Two of them are related to the particle being outside of the $x$ dimension, either $x<0$ or $x>a$. The number of such pairs is
\begin{equation}\label{}
  \Theta_{{\rm II},x}(r) = 2 \int_{0}^{a} dx \ \Omega(x) b c r^{2} n^2 = 2 \pi b c r^{3} n^2,
\end{equation}
where
\begin{equation}\label{eq:021}
  \Omega(x) = 2 \pi \left( 1 - \frac{x}{r} \right)
\end{equation}
is the solid angle of the area outside of one face. Other terms related to $y$ and $z$ axis are obtained by simply replacing $bc$ with $ac$ or $ab$.
Therefore,
\begin{equation}
  \Theta_{\rm II} = \Theta_{{\rm II},x} + \Theta_{{\rm II},y} + \Theta_{{\rm II},z} = 2 \pi (ab + ac + bc) r^3 n^2.
\end{equation}

In step III, we include back those pairs that are counted at least twice in step II. Those are the pairs where the second point is outside of the domain of two of the three axes. There are a total of 12 cases, corresponding to the 12 edges of a cuboid.
The four cases related to $x$ and $y$ are identical. Here we focus on the case where $x<0$ \& $y<0$,
\begin{equation}\label{eq:3xy}
  \frac{1}{4} \Theta_{{\rm III},xy}(r)
  = \int_{0}^{r} dx \int_{0}^{\sqrt{r^{2} - x^{2}}} dy \ c \ \Omega(\frac{x}{r},\frac{y}{r}) r^{2} n^{2} = \int_{0}^{1} dx \int_{0}^{\sqrt{1 - x^{2}}} dy ~ \Omega(x, y) ~ c ~ r^{4} n^{2}
\end{equation}
where $\Omega(x, y)$ is the solid angle of the area on a unit sphere ${\rm Sph}(x', y', z')$ with $x' > x$, $y' > y$ for positive $x$ and $y$ with $x^{2} + y^{2} < 1$.

\red{
It can be shown that the surface area of a constant latitude strip on a sphere between two longitudes is simply $R \Delta z \Delta l$, where $\Delta z$ and $\Delta l$ are the differences on the cylindrical height and on the longitudes, respectively. Using this fact, we can express the solid angle as
\begin{equation}\label{eq:022}
  \Omega(x,y) = \int_y^{\sqrt{1-x^2}} dy' ~ 2 \cos^{-1} \frac{x}{\sqrt{1 - {y'}^2}}.
\end{equation}
plugging into Eq.~(\ref{eq:3xy}) gives
\begin{align}\label{eq:202}
  \frac{\Theta_{{\rm III},xy}(r)}{4 c r^4 n^2} &= \int_0^1 dx \int_0^{\sqrt{1-x^2}} dy \int_y^{\sqrt{1-x^2}} dy' 2 \cos^{-1} \frac{x}{\sqrt{1 - y'^2}} \\
								 &= \int_0^1 dx \int_0^{\sqrt{1-x^2}} dy' \int_0^{y'} dy ~ 2 \cos^{-1}\frac{x}{\sqrt{1-y'^2}} \\
								 &= \int_0^1 dx \int_0^{\sqrt{1-x^2}} dy' ~ y' 2 \cos^{-1}\frac{x}{\sqrt{1-y'^2}} \\
								 &= \int_0^1 dy' \int_0^{\sqrt{1-y'^2}} dx ~ y' 2 \cos^{-1}\frac{x}{\sqrt{1-y'^2}} \\
								 &= \int_0^1 dy' ~ y' 2 \sqrt{1-y'^2} \\
  &= \frac{2}{3} \ ,
\end{align}
where we swapped the order of $y$ and $y'$ integration in step 1 and swapped the order of $x$ and $y'$ integration in step 3. Here we have found $\Theta_{{\rm III},xy}(r) = 8~c~r^{4} n^{2} / 3$.
Similarly, other terms related to $x$ and $z$ or to $y$ and $z$ is simply obtained by replacing $c$ with $b$ or $a$. Therefore,
\begin{equation}\label{eq:22}
  \Theta_{\rm III} = \frac{8}{3}(a + b + c) r^4 n^{2}.
\end{equation}
}

In step IV, we exclude those that are overlapped in step III. Those are the pairs where the second point is outside of the domain in all three axes. There are a total of 8 identical cases, corresponding to the 8 corners of a cuboid. Here we focus on the case where $x<0~\&~y<0~\&~z<0$.
\begin{equation}\label{eq:23}
  \frac{1}{8} \Theta_{\rm IV}(r) =
  \int_{0}^{r} dx \int_{0}^{\sqrt{r^2-x^2}} dy \int_0^{\sqrt{r^2 - x^2 - y^2}} dz \ \Omega(\frac{x}{r},\frac{y}{r}, \frac{z}{r}) \ r^2 n^2 = \int_{0}^{1} dx \int_{0}^{\sqrt{1-x^2}} dy \int_0^{\sqrt{1 - x^2 - y^2}} dz \ \Omega(x, y, z) \ r^5 n^2,
\end{equation}
where $\Omega(x,y,z)$ is the area on a unit sphere Sph($x',y',z'$) with $x' > x$, $y' > y$, $z' > z$ for positive $x$, $y$, and $z$ with $x^2+y^2+z^2 < 1$.

\red{
Following the same logic of Eq.~(\ref{eq:022}), we have
\begin{equation}
\Omega(x,y,z) = \int_z^{\sqrt{1-x^2-y^2}} dz' \left( \cos^{-1}\frac{x}{\sqrt{1-z'^2}} - \sin^{-1}\frac{y}{\sqrt{1-z'^2}} \right)
\end{equation}
Plugging into Eq.~(\ref{eq:23}) and we get
\begin{align}
  \frac{\Theta_{\rm IV}(r)}{8 r^5n^2} &=
\int_0^1 dx \int_0^{\sqrt{1-x^2}} dy \int_0^{\sqrt{1-x^2-y^2}} dz \int_z^{\sqrt{1-x^2-y^2}} dz' ~ \left( \cos^{-1}\frac{x}{\sqrt{1-z'^2}} - \sin^{-1}\frac{y}{\sqrt{1-z'^2}} \right) \\
&= \int_0^1 dx \int_0^{\sqrt{1-x^2}} dy \int_0^{\sqrt{1-x^2-y^2}} dz' \int_0^{z'} dz ~ \left( \cos^{-1}\frac{x}{\sqrt{1-z'^2}} - \sin^{-1}\frac{y}{\sqrt{1-z'^2}} \right),
\end{align}
where we have swapped the order of $z$ and $z'$ integration.
Perform the integration on $z$ and rename $z'$ to $z$ for notational simplicity
\begin{equation}
  \frac{\Theta_{\rm IV}(r)}{8 r^5n^2} =
\int_0^1 dx \int_0^{\sqrt{1-x^2}} dy \int_0^{\sqrt{1-x^2-y^2}} dz ~ z \left( \cos^{-1}\frac{x}{\sqrt{1-z^2}} - \sin^{-1}\frac{y}{\sqrt{1-z^2}} \right).
\end{equation}
Noticing that $x,y,z$ are integrated over an octant, we swap the order to put $z$ on the outside
\begin{equation}
  \frac{\Theta_{\rm IV}(r)}{8 r^5n^2} =
\int_0^1 dz \int_0^{\sqrt{1-z^2}} dx \int_0^{\sqrt{1-x^2-z^2}} dy ~z \left( \cos^{-1}\frac{x}{\sqrt{1-z^2}} - \sin^{-1}\frac{y}{\sqrt{1-z^2}} \right).
\end{equation}
Perform variable change $u = x/\sqrt{1-z^2}$, $v = y/\sqrt{1-z^2}$, and move forward
\begin{align}
  \frac{\Theta_{\rm IV}(r)}{8 r^5n^2}
  &= \int_0^1 dz \int_0^1 du \int_0^{\sqrt{1-u^2}} dv ~z (1-z^2)  (\cos^{-1}u - \sin^{-1}v) \\
&= \int_0^1 dz ~z (1-z^2) \int_0^1 du \int_0^{\sqrt{1-u^2}} dv  (\cos^{-1}u - \sin^{-1}v) \\
&= \int_0^1 dz ~z (1-z^2) \int_0^1 du ~(1 - u) \\
&= \int_0^1 dz ~z (1-z^2) ~\frac{1}{2} \\
&= \frac{1}{8}.
\end{align}
Therefore,
\begin{equation}\label{eq:26}
  \Theta_{\rm IV}(r) = r^{5} n^{2}.
\end{equation}
}

Finally, combining steps I, II, III, and IV, the total number of valid random-random pairs for the cuboidal region
with sides $a$, $b$, and $c$ and object density $n$ is given by
\begin{equation}\label{eq:30}
  \Theta(r) = \Theta_{\rm I} - \Theta_{\rm II} + \Theta_{\rm III} - \Theta_{\rm IV}
  = \left[ 4 \pi abc r^{2} - 2 \pi (ab + ac + bc) r^3 + \frac{8}{3} (a + b + c) r^4 - r^5 \right] n^2.
\end{equation}
The normalized random-random pair count is therefore
\begin{equation}\label{eq:31}
  \arc{RR}_{\rm cuboid}(r) = \frac{\Theta(r)}{(abcn)^2}
			  = \frac{4\pi}{abc} r^2 - \frac{2\pi (ab + ac + bc)}{a^2 b^2 c^2} r^3 +
  \frac{8}{3} \frac{a + b + c}{a^{2} b^{2} c^{2}} r^4 - \frac{1}{a^{2} b^{2} c^{2}} r^5,
\end{equation}
for $r \le \red{\min(a,b,c)}$.
The antiderivative is given below for an easy calculation of its integration over $r$.
\begin{equation}\label{eq:31.1}
  {\cal F}[\arc{RR}_{\rm cuboid}(r)] = \frac{4\pi}{3abc} r^3 - \frac{\pi (ab + ac + bc)}{2 a^2 b^2 c^2} r^4 + \frac{8}{15} \frac{a + b + c}{a^{2} b^{2} c^{2}} r^5 - \frac{1}{6 a^{2} b^{2} c^{2}} r^6.
\end{equation}
In a special case where $a=b=c$, Eq.~(\ref{eq:31}) becomes
\begin{equation}\label{eq:32}
  \arc{RR}_{\rm cuboid}(r) = \frac{4\pi}{a^3} r^2 - \frac{6\pi}{a^4} r^3 +
  \frac{8}{a^5} r^4 - \frac{1}{a^{6}} r^5.
\end{equation}

\subsection{Circles}

Now we consider a circle in 2D space. Without loss of generality, we assume the radius of the circle is unity.
The calculation of $\arc{RR}(r)$ is done in two steps where we take care of the whole region ignoring boundaries and then make corrections from the edge.

In step I, consider all possible pairs where the first point is inside the circle and ignore edge effects. The number of pairs separated by $r$ is given by
\begin{equation}\label{eq:40}
  \Theta_{\rm I}(r) = \pi 1^{2} \ 2\pi r \ n^2 = 2 \pi^{2} r n^{2}.
\end{equation}

In step II, we exclude the pairs where the second point is outside of the circle. The number of such pairs is given by the following integral
\begin{equation}\label{eq:41}
  \Theta_{\rm II}(r) = \int_{1-r}^1 dx \ 2 \pi x \phi(x) r n^2,
\end{equation}
where
\red{
\begin{equation}\label{eq:42}
  \phi(x) = 2 \cos^{-1} \frac{1 - x^{2} - r^{2}}{2 x r}
\end{equation}
}
is the angle of the arc outside of the circle (Fig.~\ref{fig:1.1}, right). Substituting it into Eq.~(\ref{eq:41}) gives
\red{
\begin{equation}\label{eq:43}
  \Theta_{\rm II}(r) = \left(\pi r^{2}\sqrt{4-r^{2}} + 4 \pi r \sin^{-1} \frac{r}{2} \right) n^{2}
\end{equation}
}

Combining steps I and II, the total number of valid random-random pairs for a unitary circular region with object density $n$ is given by
\begin{equation}\label{eq:44}
  \Theta(r) = \Theta_{\rm I}(r) - \Theta_{\rm II}(r) = \red{ \left( 2 \pi^2 r - \pi r^{2} \sqrt{4 - r^{2}} - 4 \pi r \sin^{-1}\frac{r}{2} \right) n^2, }
\end{equation}
and the normalized random-random pair count,
\begin{equation}\label{eq:46}
  \arc{RR}_{\rm circ}(r) = \frac{\Theta(r)}{(\pi \cdot 1^{2} \cdot n)^2} = \red{ 2 r - \frac{1}{\pi} r^2 \sqrt{4 - r^{2}} - \frac{4}{\pi} r \sin^{-1} \frac{r}{2}, }
\end{equation}
for $r \le 1$. \red{The antiderivative is given below for an easy calculation of its integration over $r$.
\begin{equation}\label{eq:45.2}
{\cal F}[\arc{RR}_{\rm circ}(r)] = r^2-\frac{1}{4\pi} \sqrt{4-r^2} \left(r^2+2\right) r + \frac{2}{\pi} \left(1-r^2\right) \sin^{-1} \frac{r}{2}.
\end{equation}
}

\subsection{Spheres}

In the final one of the four geometries, we consider a unit sphere. We calculate $\arc{RR}(r)$ in two steps where we take care of the whole region ignoring boundaries and then make corrections from its edge.

In step I, we consider all possible pairs where the first point is inside the circle and ignore edge effects. The number of pairs separated by $r$ is given by
\begin{equation}\label{eq:50}
  \Theta_{\rm I}(r) = \frac{4\pi}{3} 1^3 4\pi r^2 n^2 = \frac{16 \pi^2}{3} r^2 n^2.
\end{equation}

In step II, we exclude the pairs where the second point is outside of the unit sphere. The number of such pairs is
\begin{equation}\label{eq:51}
  \Theta_{\rm II}(r) = \int_{1-r}^1 dx \ 4\pi x^2 \Omega(x) r^2 n^2 = \left(4 \pi^{2} r^3 - \frac{\pi^{2}}{3} r^5\right) n^2,
\end{equation}
where
\begin{equation}\label{eq:52}
 \Omega(x) = \left( 1 + \frac{x^2 + r^2 - 1}{2 x r}\right) 2 \pi
\end{equation}
is the solid angle of the area on a sphere of radius $r$ that is outside of the unit sphere.

Combining step I and step II, the total number of valid random-random pairs for the unitary spherical region with object density $n$ is given by
\begin{equation}\label{eq:53}
  \Theta(r) = \Theta_{\rm I} - \Theta_{\rm II} = \left( \frac{16 \pi^2}{3} r^{2} - 4\pi^{2} r^{3} + \frac{\pi^3}{3} r^5 \right) n^2,
\end{equation}
and the normalized random-random pair count,
\begin{equation}\label{eq:53.2}
  \arc{RR}_{\rm sph}(r) = \frac{\Theta(r)}{\left( \frac{4\pi}{3} n \right)^2} = 3 r^2 - \frac{9}{4} r^3 + \frac{3}{16} r^5,
\end{equation}
for $r \le 1$. The antiderivative is given below for an easy calculation of its integration over $r$.
\begin{equation}\label{eq:54}
  {\cal F}[\arc{RR}_{\rm sph}(r)] = r^3 - \frac{9}{16} r^4 + \frac{1}{32} r^6.
\end{equation}

\section{Accounting for Edge Corrections in $\arc{RR}$}
\label{sec:dis}

\begin{table}
  \centering
  \caption{\label{tab:1} \red{Percentage errors} $\epsilon(\hat{r})$ of the calculated $\arc{RR}(\hat{r})$ when certain edge corrections are not included.
	It is shown that all corrections in the rectangular, circular, and spherical cases must be considered in order to limit the relative error to sub-percent. For the cuboidal case, however, the corner correction causes at most sub-percent changes (highlighted) to the calculated $\arc{RR}$ at all radii. }
\begin{tabular}{lllllll}
  \toprule
 {\bf Field geometry} & {\bf Corrections included} & $\epsilon(0.001)$ & $\epsilon(0.01)$ & $\epsilon(0.1)$ & $\epsilon(0.2)$ & $\epsilon(0.4)$ \\
  \hline
Rectangle & None & 0.13    &  1.3     & 14      & 32     &  85     \\
& Edge & -3.2e-05 & -0.0032  & -0.36   & -1.7   &  -9.4   \\
& Edge + Corner & 0 & 0 & 0 & 0 & 0 \\
Cuboid & None & 0.15    &  1.5     & 17      & 38     &   100 \\
& Face & -6.4e-05 & -0.0065  & -0.73   & -3.4   & -19     \\
& Face + edge & {\bf 8e-09}   &  {\bf 8.1e-06} &  {\bf 0.0093} &  {\bf 0.088} &   {\bf 1.0}     \\
& Face + edge + corner & 0 & 0 & 0 & 0 & 0 \\
Circle & None & 0.064   &  0.64    &  6.8    & 15     &  34     \\
  & Edge & 0 & 0& 0& 0& 0 \\
Sphere & None & 0.075   &  0.76    &  8.1    & 18     &  42     \\
  & Edge & 0 & 0 & 0 & 0 & 0 \\
\hline
\end{tabular}
\end{table}

The formulas of $\arc{RR}(r)$ derived in this work, Eqs.~(\ref{eq:1.9}), (\ref{eq:31}), (\ref{eq:46}), and (\ref{eq:53.2}), have included all edge corrections and are precise.
However, the calculation of $\arc{DR}(r)$ is more complex because iterations over all data points is required. Given the complexity of the edge-correction formulas, it would be beneficial if some of the terms could be ignored without sacrificing accuracy.

To account for the contributions from various edge effects, we break the formulas of $\arc{RR}$, Eqs.~(\ref{eq:1.9}), (\ref{eq:31}), (\ref{eq:46}), and (\ref{eq:53.2}), into various terms based on the powers of $r$ and rewrite them as follow:
\begin{align}
  &\arc{RR}_{\rm rect}(r) = \frac{2\pi}{ab} r \left(1 - \frac{2(a+b)}{\pi a b}r + \frac{1}{\pi a b} r^2 \right),\\
  &\arc{RR}_{\rm cuboid}(r) = \frac{4\pi}{abc} r^2 \left(1 - \frac{ab + ac + bc}{2 a b c}r +
  \frac{2}{3 \pi} \frac{a + b + c}{a b c} r^2 - \frac{1}{4\pi a b c} r^3 \right),\\
  &\arc{RR}_{\rm circ}(r) = 2 r \left(1 - \frac{2}{\pi} r + \frac{1}{12\pi} r^3 + \frac{1}{320 \pi} r^{5} + O(r^7) \right),\\
  &\arc{RR}_{\rm sph}(r) = 3 r^2 \left(1 - \frac{3}{4} r + \frac{1}{16} r^3 \right).
\end{align}
The 0th, 1st and 2nd-order terms of $\arc{RR}_{\rm rect}$ correspond to the inside, the edges, and the corners of the rectangle.
The 0th, 1st, 2nd, and 3rd-order terms of $\arc{RR}_{\rm cuboid}$ are the inside, the faces, the edges, and the corners of the cuboid.
The 0th-order term of $\arc{RR}_{\rm circ}$ or $\arc{RR}_{\rm sph}$ is the inside of the circle or sphere and the other terms are edge correction.

$\arc{RR}$ are calculated up to various degrees of edge corrections and are compared to its precise values. The fractional errors at a list of radii are listed in Table~\ref{tab:1}. Of all the geometries, unitary sides or radius is assumed.
We conclude that in all but the cuboidal case, all the edge-correction terms are necessary in order to limit the errors of $\arc{RR}(r)$ to within $1\%$ at all $r$. In the cuboidal case, the error caused by ignoring the corner correction is sub-percent even at large $r$. This latter fact helps to significantly simplify the computation of $\arc{DR}_{\rm rect}$ in the following section.

\section{Computing $\arc{DR}$ Analytically in $O(N_{\rm g})$ Time}
\label{sec:dr}

The computation of $\arc{DR}$ requires iterations over all data particles. For a data point away from the edges of the survey area, the number of data-random pairs shared by this data point is simply $2 n \pi r dr$ in 2D case or $4 n \pi r^2 dr$ for 3D case, where $n$ is the mean number density of the random catalogue. When a data point is close to the edges, corrections are necessary to exclude pairs where the random point is outside.

In the previous section, we have derived various edge-correction terms for the four geometries. In this section, we use those terms to calculate the contribution of each data point to the data-random pair counts with full edge corrections. We work out extra steps of integrating over $r$ for an easy calculation of the edge corrections without numerical integration. The subsequent algorithms to compute the edge-corrected $\arc{DR}(r)$ in $O(N_{\rm g})$ time is presented in Appendix~\ref{sec:A}.

\subsection{Rectangles}

For a data point $D$ that is close to any of the edges of a rectangle, a circle around $D$ with radius $r$ has a part outside of the region when $r > y$, where $y$ is the distance from $D$ to the edge. The angle of this arc, $\phi(y)$, is given by Eq.~(\ref{eq:1.2.2}). The integration of the length of the arc, $\phi(r)r$, over $r$ is therefore
\begin{equation}\label{eq:a2.1}
 {\cal F}(r; y) = \int \phi(y) r \, dr = \int 2 \cos^{-1}\left( \frac{y}{r} \right) r \, dr = r^2 \cos^{-1} \frac{y}{r} - y \sqrt{r^2 - y^2}.
\end{equation}

When $D$ is close to a corner, i.e., $\sqrt{x^2 + y^2} < r$, where $x$ and $y$ are the distances to the two sides, the part of the arc outside of both extended sides are excluded twice in the previous step (see Fig.~\ref{fig:1.1}, {\it Center}), therefore we need to include them back. The angle of this arc, $\phi(x,y)$, is given by Eq.~(\ref{eq:1.5}). The integration of the length of the arc, $\phi(r)r$, over $r$ is therefore
\begin{align}\label{eq:a2.2}
  {\cal F}(r; x, y) = \int \phi(x, y) r dr
  = \frac{1}{4} \left[ \pi r^2 - 2 x \sqrt{r^2 - x^2} - 2 y \sqrt{r^2 - y^2} - 2 r^2 \left( \sin^{-1} \frac{x}{r} + \sin^{-1} \frac{y}{r} \right) \right] + C
\end{align}
where $C = x y$ is chosen such that ${\cal F}(r; x, y)$ approaches 0 when $r \rightarrow \sqrt{x^2 + y^2}$.

Based on Eq.~(\ref{eq:a2.1}) and Eq.~(\ref{eq:a2.2}), we write an algorithm to compute the data-random pair counts analytically by doing iterations over all data points where the contribution from each data point is calculated 1) as $2\pi r \, dr$ if not touching the edges, or 2) using Eq.~(\ref{eq:a2.1}) and Eq.~(\ref{eq:a2.2}) it touches at least one of the edges. This algorithm has a time complexity of $O(N_{\rm g})$, where $N_{\rm g}$ is the number of galaxies or particles. The pseudocode is presented in Appendix~\ref{sec:A}.

\subsection{Cuboids}
\label{sec:drcub}

The calculation of Data-Random pair counts in a cuboidal region is similar to that in a rectangular region, with the angle $\phi$ replaced by a solid angle $\Omega$, plus an extra step to take care of the corners of the cuboid.

For a data point $D$ that is close to any of the faces of a cuboid, the part on a sphere centered at $D$ with radius $r$ that is outside of the cuboid is given by Eq.~(\ref{eq:021}), where $x$ is the distance to the edge. The integration of the area of this part, $\Omega(x) r^2$, over $r$ is therefore
\begin{equation}\label{eq:a2.3}
  {\cal F}(r; x) = \int \Omega(x) r^{2} dr = \int 2 \pi (r^2 - xr) dr = 2 \pi \left( \frac{r^3}{3} - \frac{x r^2}{2} \right).
\end{equation}

When $D$ is close to one of the edges of the cuboid such that $\sqrt{x^2 + y^{2}} < r$, where $x$ and $y$ are the distances to two adjacent faces, the part of the sphere outside of both faces are excluded twice in the previous step, therefore we need to include them back. The solid angle of this surface, \red{when assuming $r=1$, is given by Eq.~(\ref{eq:022}), which equals
\begin{equation}\label{eq:whatever}
  \Omega(x,y) = \left( \frac{1}{2} - x -y \right)\pi + 2 x \tan^{-1} \frac{y}{t} + 2 y \tan^{-1} \frac{x}{t} + \tan^{-1} \frac{t^{2} - x^{2}y^2}{2 x y t},
\end{equation}
where $t \equiv \sqrt{1 - x^2 - y^2} $.
Replacing $x$ with $x/r$ and $y$ with $y/r$, we get $\Omega(r; x, y)$.} The integration of the area of the surface, $\Omega(r;x,y)r^2$, over $r$ is therefore
\begin{align}
  {\cal F}(r; x, y) = &\int \Omega(r;x,y) r^2 dr \nonumber \\
  = &\int \left[ \pi \left(\frac{r^2}{2}-r x-r y\right)+2 r y \tan ^{-1}\left(\frac{x}{h}\right)+2 r x \tan ^{-1}\left(\frac{y}{h}\right) + r^2 \tan^{-1} \frac{r^2 h^{2} - x^{2}y^2}{2 x y r h} \right] dr \nonumber \\
  = & \ r^2 \left(y \tan ^{-1}\left(\frac{x}{h}\right)+x \tan ^{-1}\left(\frac{y}{h}\right)\right)+\frac{1}{6} \pi  r^2 (r-3 (x+y)) \nonumber \\
  & +x^3 \cot ^{-1}\left(\frac{y}{h}\right)+y^3 \cot ^{-1}\left(\frac{x}{h}\right) + 2 x y h \nonumber \\
  & + \frac{x^{3}}{3}\left[\tan^{-1}\left(\frac{r x+x^{2}+y^{2}}{hy}\right) + \tan^{-1}\left(\frac{-r x+x^{2}+y^{2}}{h y}\right)\right] \nonumber \\
  & + \frac{y^{3}}{3}\left[\tan^{-1}\left(\frac{r y+x^{2}+y^{2}}{hx}\right) + \tan^{-1}\left(\frac{-r y+x^{2}+y^{2}}{h x}\right) \right] \nonumber \\
  & - \frac{r^{3}}{3} \tan^{-1}\left(\frac{x^{2} y^{2} - r^{2} h^2}{2 x y rh}\right) \nonumber \\
  & - \frac{4}{3}yxh + C \label{eq:a2.4}
\end{align}
where $h \equiv \sqrt{r^2 - x^2 - y^2} $ and $C = - \frac{\pi}{3}(x^3 + y^3)$ is chosen such that $\lim_{r \rightarrow \sqrt{x^2 + y^2}}{\cal F}(r; x, y) = 0$, or $\lim_{h \rightarrow 0^+}{\cal F}(r; x, y) = 0$.

To compute $\arc{DR}(r)$ with perfect accuracy, one would need to take into account the effects of the corners, which is when a sphere around $D$ is outside of three adjacent faces simultaneously. They are included back twice in the previous step and need to be subtracted again.
However, this effect is ignored in this work because \red{the computation is too complicated and} its contribution is at
most sub-percent (Table~\ref{tab:1}).

An algorithm to compute $\arc{DR}(r)$ analytically is presented in
Appendix~\ref{sec:A}.

\subsection{Circles}

When a data point $D$ is close to the edge of a unitary circular survey region, a circle around $D$ with radius $r$ has a segment outside of the region when $r > 1 - x$, where $x$ is this point's distance to the regional center. The angle of this arc is given by Eq.~(\ref{eq:42}). The integration of  the length of the arc, $\phi(r)r$, over $r$ is therefore
\red{
\begin{align}\label{eq:a3.1}
  {\cal F}(r; x) = \int \phi(x) r dr = \frac{\eta }{2} + r^2 \cos ^{-1}\left(\frac{1-r^2-x^2}{2 x r}\right)+\sin^{-1}\left(\frac{1-r^2+x^2}{2x }\right) + C,
\end{align}
}
where $\eta \equiv \sqrt{(-r+x+1) (r-x+1) (r+x-1) (r+x+1)}$ and $C = -\pi / 2$ is chosen such that $\lim_{r \rightarrow (1 - x)^+}{\cal F}(r) = 0$.

An algorithm to compute $\arc{DR}(r)$ analytically is presented in Appendix~\ref{sec:A}.

\subsection{Spheres}

When a data point $D$ is close to the edge of a unitary spherical region, a circle around $D$ with radius $r$ has a segment outside of the region when $r > 1 - x$, where $x$ is this point's distance to the center of the region. The solid angle of this segment is given by Eq.~(\ref{eq:52}). The integration of the surface area of this segment, $\Omega(r)r^2$, over $r$ is therefore
\begin{align} \label{eq:a3.2}
  {\cal F}(r; x) = \int \Omega(x) r^{2} dr = \frac{\pi  r^2 \left(-6+3 r^2+8 r x+6 x^2\right)}{12 x}.
\end{align}

An algorithm to compute $\arc{DR}(r)$ analytically in $O(n)$ time is given in Appendix~\ref{sec:A}.

\section{Comparisons with the Monte Carlo Method}
\label{sec:3}

\begin{figure*}[ht]
  \centering
  \includegraphics[scale=0.9]{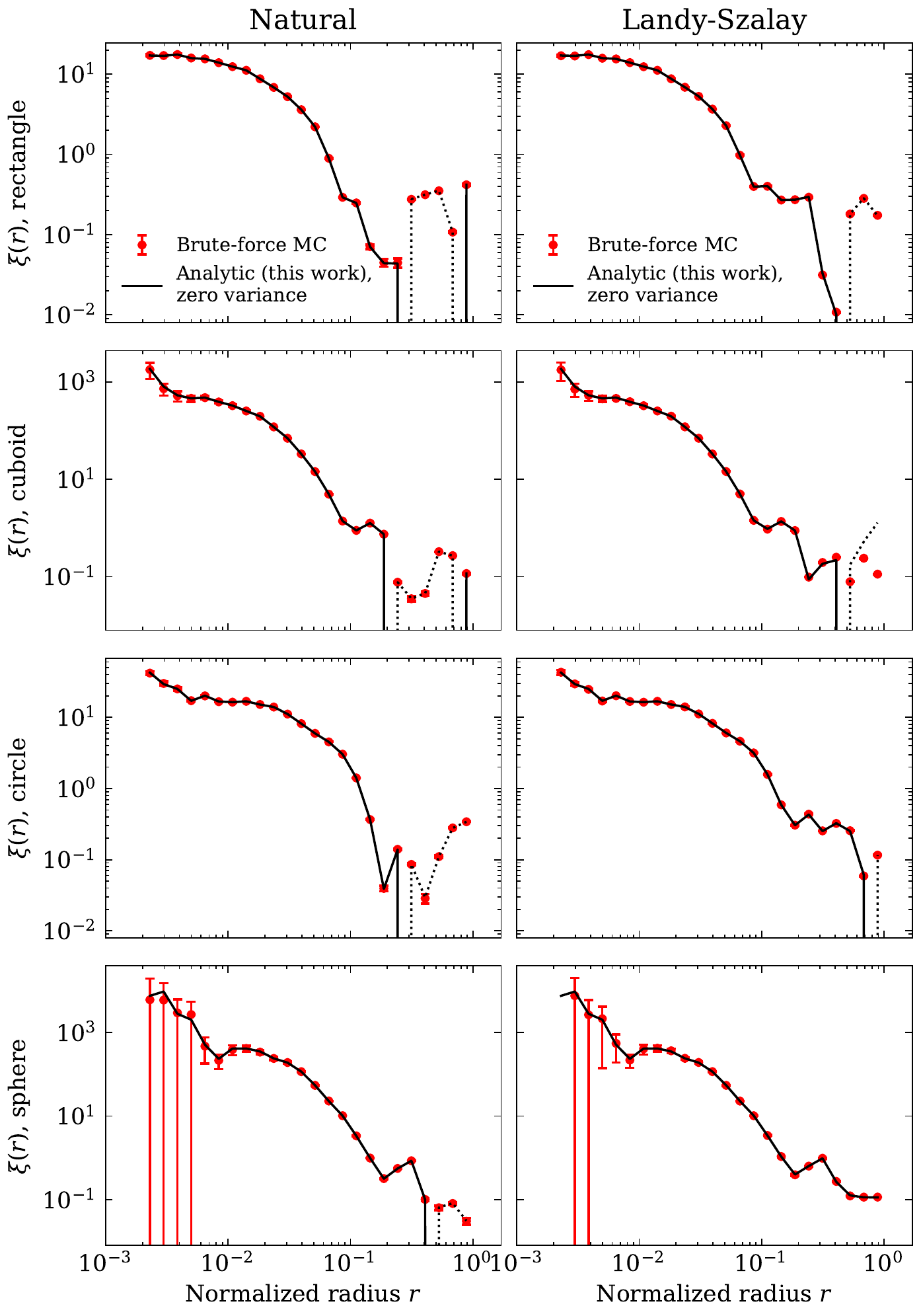}
  \caption{\label{fig:1} Comparing the two-point correlation functions $\xi(r)$ of a mock galaxy population calculated from the analytic method from this work (solid black curve) with that from \red{brute-force} Monte Carlo method (red dots and error bars).
	The former is shown to be the asymptotic limit of the latter as the size of the random catalogue goes to infinity, evident from the fact that the means (red dots) of the brute-force Monte Carlo estimations strictly follow the analytic predictions.
	The natural estimator, $\xi_{0} = \arc{DD}/\arc{RR} - 1$, and the Landy-Szalay estimator, $\xi_{\rm LS} = \arc{DD}/\arc{RR} - 2 \arc{DR}/\arc{RR} + 1$, are applied in the left and right panel, respectively.
	The error bars are the standard deviations estimated from 40 random catalouges. Negative $\xi$ is donated as dotted lines.
	While the Monte Carlo estimations exhibit significant scattering at small $r$, the analytic method has zero variance at all scales.
  }
\end{figure*}

In this section, we compare the accuracy and speed of the analytic method from this work
with the Monte Carlo method used in most literature.

The test data set we use is a mock galaxy catalogue inside a
$100~{\rm cMpc}$ cube from the
\texttt{RefL0100N1504\_Subhalo} simulation of the EAGLE database \citep{McAlpine2016}.
The positions of the galaxies are normalized to a unitary box for simplicity.
For the 2D geometries, the third dimension of the positions is removed.
For the circular or spherical cases, the box is shifted, normalized, and trimmed to a unit sphere centered at the origin.

We compare the $\xi(r)$ computed using our analytic method with that using a brute-force Monte Carlo code\footnote{The code we are using is {\tt scipy.spatial.cKDTree}, which is supposed to embrace a tree-based algorithm and have better performance than brute force. However, what we observe is that both its performance and accuracy under default setting are very close to brute force with a scaling close to $O(N^2)$.} (Fig.~\ref{fig:1}), using both the Natural estimator (left column) and the Landy-Szalay estimator (right column), applying to the four geometries we have discussed.
For a quick runtime, we choose 1000 galaxies from the galaxy catalogue and adopt a low random-to-data ratio 16 to make the computation manageable on a laptop.
A total of 40 random catalogues are generated to estimate the mean and standard deviation of $\arc{RR}(r)$ and $\arc{DR}(r)$ in 25 separation bins, evenly distributed in logarithmic scale. They are then passed to $\xi(r)$ to estimate its mean (red dots) and standard deviation (errorbars).
Our analytic calculation is, by construction, the asymptotic limit of the Monte Carlo calculation as $N_{\rm r} \rightarrow \infty $, hence zero variance.

We observe that the Monte Carlo estimations of $\xi(r)$ have means strictly following the analytic results \footnote{The only exception is $\xi_{\rm cuboid}(r)$ at $r \ge 0.4$. This is caused by the exclusion of corner corrections (Section~\ref{sec:drcub}).}.
The perfect agreement with the Monte Carlo estimations demonstrates the validity of the analytic formulas of $\arc{RR}$ and the analytic algorithms of $\arc{DR}$ proposed in this work.
While the brute-force Monte Carlo estimations exhibit significant scattering at small scales, the analytic method has zero variance at all scales.

\begin{figure}
  \centering
  \includegraphics[]{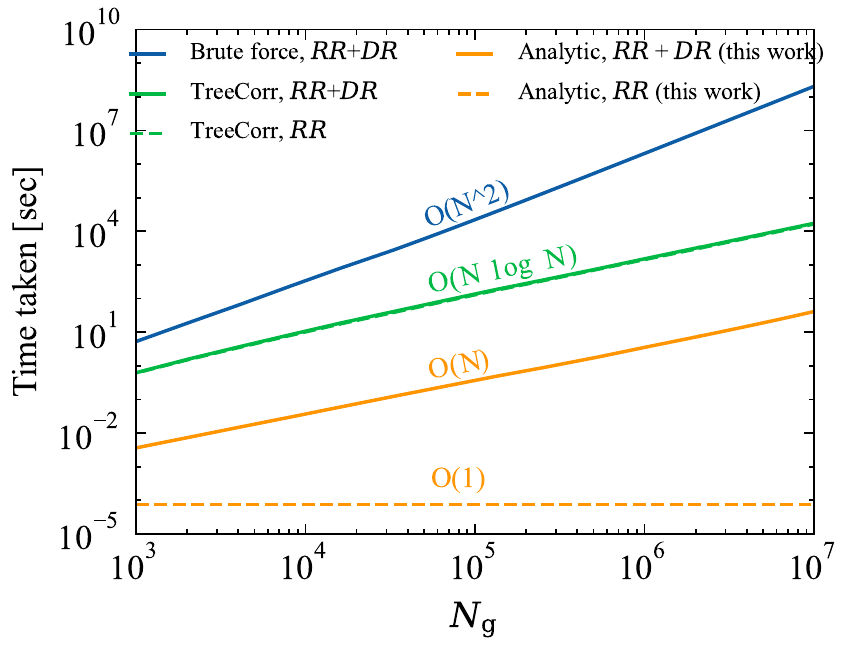}
  \caption{\label{fig:c1}\red{Comparing the speed of the analytic approach from this work to that of the traditional Monte Carlo method.
	  The performance of our analytic method is 3 to 6 orders of magnitude higher than the brute force method and 2.5 orders of magnitude higher than a fast tree-based algorithm for sizes of galaxy catalogues explored.
  }
  }
\end{figure}

\red{In practice, the brute-force Monte Carlo method is usually used for accurate computation of the TPCF of a small galaxy catalogue. For large galaxy catagloues, people tend to faster tree-based algorithms with some sacrifice in accuracy.}
We compare the time it takes to compute \red{$\arc{RR}(r)$ and $\arc{DR}(r)$} using different methods (Fig.~\ref{fig:c1}). \red{The brute force Monte Carlo code is running with a random catalogue 50 times bigger than the data catalogue, a typical practice in the community. \textsc{TreeCorr} \citep{Jarvis2004} is used as a representative of tree-based code. Our method does not use random catalogues since all terms are calculated analytically. Despite running with Python, all the programs do the actual computation either with C/C++ or using {\sc Numba} to achieve comparable speed to C.
  Our analytic method computes $RR+DR$ at a speed 3 to 6 orders of magnitude faster than brute-force Monte Carlo method and 2.5 orders of magnitude faster than tree-based code\footnote{The benchmarking of \textsc{TreeCorr} is performed under the default setting ({\tt bin\_slop=1}). For a better accuracy, say \texttt{bin\_slop=0.1}, the computation time of \textsc{TreeCorr} is supposed to be $\sim 30$ times longer (Siewert et al. 2020).} for galaxy catalogues with sizes up to 10 million.
  At $N_{\rm g} = 10^7$, while it takes the brute force Monte Carlo code a projected time of 100,000 hours and the tree-based code 6 hours, our analytic method gets it done in under 1 minute on a single core.
  Particularly, the calculation of $\arc{RR}$ through our analytic method takes only $10^{-4}$ second, independent of the size of the galaxy catalogue.
}

\section{Summary and Discussion}
\label{sec:sum}

We have proposed a new analytic method to compute the galaxy TPCF in an extremely efficient and accurate way, \red{applicable to mask-free surveys with regular geometries.}
\red{
We have derived simple closed-form formulas of the normalized random-random pair counts $\arc{RR}(r)$ for mask-free surveys with the following geometries: rectangles with sides $a$ and $b$ (Eqs.~(\ref{eq:1.9}) and (\ref{eq.1.9v2})), cuboids with sides $a$, $b$, and $c$ (Eq.~(\ref{eq:31})), a unit circle (Eq.~(\ref{eq:46})), and a unit sphere (Eq.~(\ref{eq:53.2})). With all edge corrections fully considered, these formulas calculate $\arc{RR}(r)$ with perfect accuracy and zero variance in $O(1)$ time.
We have also presented a set of pseudocode to compute the data-random pair counts $\arc{DR}(r)$ analytically and precisely with zero variance in $O(N_{\rm g})$ time, applicable to the above-mentioned geometries. These algorithms are presented in Appendix~\ref{sec:A} with a link to the Python code.}
$\arc{RR}(r)$ and $\arc{DR}(r)$ together can be used to calculate $\xi(r)$ using any estimator.

\red{
  We have applied our method to a mock galaxy catalogue from the EAGLE simulation \citep{McAlpine2016} and compared the calculated $\xi(r)$ with that from the brute-force Monte Carlo method (Fig.~\ref{fig:1}). Perfect agreement is found.
  We have also compared the speed of our method with that of the brute-force Monte Carlo method, which has $O(N^2)$ scaling, and the tree-based method, which has $O(N \log N)$ scaling.
  Our analytic method is 3 to 6 orders of magnitude faster than the brute force Monte Carlo method and 2.5 orders of magnitude faster than the tree-based code.}
Our proposed method is favored over the traditional numerical method whenever applicable.

\red{Our method can be used to replace the brute-force Monte Carlo method as the benchmark of evaluating the accuracy of existing or new code. It could also be particularly useful in the study of non-linear correlation functions and power spectra of galaxies, dark matters, and haloes with exceptional accuracy and efficiency. While most cosmological simulations are done with periodic boundary conditions where the correlation is trivially analytic, researchers may find its applications in particular cases: the study of local clustering of a subregion with high precision, or the study of isolated systems.}

\red{
For a realistic survey with masks, weight patterns, and/or irregular shapes, the applicability of our approach is very limited, although in special circumstances some applications can be found. Our method can be used to account for masks with regular shapes on top of a mask-free background. Clustering in subregions of a large survey can be explored with ease and with high precision. Although, in both cases, the correlation between the mask or weighted area and the background has to be computed numerically.
}

Our proposed analytic method is also directly applicable to galactic angular TPCFs when the survey area is close to Euclidean (flat).

\begin{acknowledgments}
  We are grateful to the referee for providing constructive suggestions to improve the paper.
\end{acknowledgments}

\bibliographystyle{apj}
\bibliography{BIB_TPCF}

\appendix

\section{Algorithms to Compute $\arc{DR}$ Analytically in $O(N_{\rm g})$ Time}
\label{sec:A}

\renewcommand\thefigure{\thesection.\arabic{figure}}
\setcounter{figure}{0}

\begin{figure}
  \caption{\label{alg1}An algorithm for precise calculations of $\arc{DR}(r)$ in $O(N_{\rm
  g})$ time, applying to rectangular regions.}
\hrule
\begin{algorithmic}[1]
	\State $a, b$: dimensions of the rectangular region.
	\State $N$: the total number of data points.
	\State $rlist$: array of boundaries defining the real space radial bins in which pairs are counted.
	\State $rsteps$: a list of discrete differences along $rlist$.
	\For {$i \in 1:(len(rlist)-1)$}
		\State $drpair \gets 0$
		\State $rthis \gets rlist[i]$
		\State $rnext \gets rlist[i+1]$
		\State $r \gets rnext$
		\For {$par \in$ data set}
			\State
			\State $drpair \gets drpair + pi * (rnext^2 - rthis^2)$ \Comment {All the
			pairs, assuming a periodic boundary condition.}
			\State
			\State $x, y$: coordinates of $par$
			\If {$x < r$}
				\State {$xgapl \gets x$}
				\Else
				\State {$xgapl \gets -1$}
				\EndIf
			\If {$x > a - r$}
				\State $xgapr \gets a - x$
			\Else
				\State $xgapr \gets -1$
			\EndIf
			\If {$y < r$}
				\State {$ygapl \gets y$}
				\Else
				\State {$ygapl \gets -1$}
				\EndIf
			\If {$y > b - r$}
				\State $ygapr \gets b - y$
			\Else
				\State $ygapr \gets -1$
			\EndIf

			\State \Comment{Exclude the edges}

			\For {$igap \in [xgapl, xgapr, ygapl, ygapr]$}
				\If {$igap > 0$}
					\If {$igap > rthis$}
						\State $F1 \gets 0$
					\Else
						\State $F1 \gets int\_rec\_edge(rthis, igap)$ \Comment {Eq.~(\ref{eq:a2.1})}
					\EndIf
					\State $F2 \gets int\_rec\_edge(rnext, igap)$
					\State $drpair \gets drpair - (F2 - F1)$
				\EndIf
			\EndFor

			\State \Comment{Include the corners back}
			\For {$(xgap, ygap) \in [(xgapl, ygapl), (xgapl, ygapr), (xgapr, ygapl), (xgapr, ygapr)]$}
			\If {$xgap > 0 \And ygap > 0 \And xgap^2 + ygap^2 < r^2$}
				\If {$xgap^2 + ygap^2 \ge rthis^2$}
					\State $F1 \gets 0$
				\Else
					\State $F1 \gets int\_rec\_corner(rthis, xgap, ygap)$ \Comment {Eq.~(\ref{eq:a2.2})}
				\EndIf
				\State $F2 \gets int\_rec\_corner(rnext, xgap, ygap)$
				\State $drpair \gets drpair + F2 - F1$
			\EndIf
			\EndFor

			\State
			\State $\arc{DR}[i] \gets drpair / (N * a * b)$
		\EndFor
	\EndFor
\end{algorithmic}
\hrule
\end{figure}

\begin{figure}
  \caption{\label{alg2}An algorithm for precise calculations of $\arc{DR}(r)$ in $O(N_{\rm g})$ time, applying to cuboidal regions.}
\hrule
\begin{algorithmic}[1]
\State $a, b, c$: dimensions of the cuboidal region.
\State $N$: total number of data points.
\State $rlist$: array of boundaries defining the real space radial bins in which pairs are counted.
\State $rsteps$: a list of discrete differences along $rlist$.
\For {$i \in 1:(len(rlist)-1)$}
  \State $drpair \gets 0$
  \State $rthis \gets rlist[i]$
  \State $rnext \gets rlist[i+1]$
  \State $r \gets rnext$
  \For {$par \in$ data set}
	\State $drpair \gets drpair + \frac{4}{3} * \pi * (rnext^3 - rthis^3)$
	\State $x, y, z$: coordinates of $par$
			\If {$x < r$}\Comment{Calculate the gaps between the particle and the edges}
				\State {$xgapl \gets x$}
				\Else
				\State {$xgapl \gets -1$}
				\EndIf
			\If {$x > a - r$}
				\State $xgapr \gets a - x$
			\Else
				\State $xgapr \gets -1$
			\EndIf
			\If {$y < r$}
				\State {$ygapl \gets y$}
				\Else
				\State {$ygapl \gets -1$}
				\EndIf
			\If {$y > b - r$}
				\State $ygapr \gets b - y$
			\Else
				\State $ygapr \gets -1$
			\EndIf
			\If {$z < r$}
				\State {$zgapl \gets z$}
				\Else
				\State {$zgapl \gets -1$}
				\EndIf
			\If {$z > c - r$}
				\State $zgapr \gets c - z$
			\Else
				\State $zgapr \gets -1$
			\EndIf

	\For {$igap \in [xgapl, xgapr, ygapl, ygapr, zgapl, zgapr]$} \Comment{Exclude the faces}
		\If {$igap > 0$}
			\If {$igap > rthis$}
				\State $F1 \gets intface(igap, igap)$ \Comment{Eq.~(\ref{eq:a2.3})}
			\Else
				\State $F1 \gets intface(rthis, igap)$
			\EndIf
			\State $F2 \gets intface(rnext, igap)$
			\State $drpair \gets drpair - (F2 - F1)$
		\EndIf
	\EndFor

	\State $xgaps \gets (xgapl, xgapr)$\Comment{Include back the edges}
	\State $ygaps \gets (ygapl, ygapr)$
	\State $zgaps \gets (zgapl, zgapr)$
            \For {$(igaps, jgaps) \in [(xgaps, ygaps), (xgaps, zgaps), (ygaps, zgaps)]$}
                \For {$gapi \in igaps$}
                    \For {$gapj \in jgaps$}\Comment{$3\times 2 \times 2 = 12$ is the number of edges in a cuboid.}
		\If {$gapi > 0 \And gapj > 0 \And gapi^2 + gapj^2 < r^2$}
			\If {$gapi^2 + gapj^2 \ge rthis^2$}
				\State $F1 \gets 0$
			\Else
				\State $F1 \gets intedge(rthis, gapi, gapj)$ \Comment {Eq.~(\ref{eq:a2.4})}
			\EndIf
			\State $F2 \gets intedge(rnext, gapi, gapj)$
			\State $drpair \gets drpair + F2 - F1$
		\EndIf
	\EndFor
	\EndFor
	\EndFor
  \EndFor
  \State $\arc{DR}[i] \gets drpair / (N * a * b * c)$ \Comment{Calculate the normalized data-random pair counts}
\EndFor
\end{algorithmic}
\hrule
\end{figure}

\begin{figure}
  \caption{\label{alg3}An algorithm for precise calculations of $\arc{DR}(r)$ in $O(N_{\rm
  g})$ time, applying to circular regions.}
\hrule
\begin{algorithmic}[1]
	\State $N$: the total number of data points.
	\State $rlist$: array of boundaries defining the real space radial bins in which pairs are counted.
	\State $rsteps$: a list of discrete differences along $rlist$.
  \For {$i \in 1:(len(rlist)-1)$}
  \State $drpair \gets 0$
  \State $rthis \gets rlist[i]$
  \State $rnext \gets rlist[i+1]$
  \State $r \gets rnext$
  \For {$par \in$ data set}
  \State
  \State $drpair \gets drpair + \pi * (rnext^2 - rthis^2)$
  \State
  \State $x, y$: coordinates of $par$
  \State $d \gets \sqrt{x^2+y^2}$
  \State $gap \gets 1 - d$

  \If {$r > gap$}
  \If {$rthis \le gap$}
  \State $F1 \gets 0.$
  \Else
  \State $F1 \gets int\_unit\_circle\_edge(rnext, d)$ \Comment{Eq.~(\ref{eq:a3.1})}
  \EndIf
  \State $F2 \gets int\_unit\_circle\_edge(rnext, d)$
  \State $drpair \gets drpair - (F2 - F1)$
  \EndIf
  \EndFor

  \State
  \State $\arc{DR}[i] \gets drpair / (N * \pi)$
  \EndFor
\end{algorithmic}
\hrule
\end{figure}

\begin{figure}
  \caption{\label{alg4}An algorithm for precise calculations of $\arc{DR}(r)$ in $O(N_{\rm g})$ time, applying to spherical regions.}
\hrule
\begin{algorithmic}[1]
	\State $N$: the total number of data points.
	\State $rlist$: array of boundaries defining the real space radial bins in which pairs are counted.
	\State $rsteps$: a list of discrete differences along $rlist$.
  \For {$i \in 1:(len(rlist)-1)$}
  \State $drpair \gets 0$
  \State $rthis \gets rlist[i]$
  \State $rnext \gets rlist[i+1]$
  \State $r \gets rnext$
  \For {$par \in$ data set}
  \State $drpair \gets drpair + \frac{4}{3} \pi * (rnext^3 - rthis^3)$
  \State $x, y, z$: coordinates of $par$
  \State $d \gets \sqrt{x^2+y^2+z^3}$
  \State $gap \gets 1 - d$
  \If {$r > gap$}
  \If {$rthis \le gap$}
  \State $F1 \gets int\_unit\_sphere\_edge(gap, d)$ \Comment{Eq.~(\ref{eq:a3.2})}
  \Else
  \State $F1 \gets int\_unit\_sphere\_edge(rthis, d)$
  \EndIf
  \State $F2 \gets int\_unit\_sphere\_edge(rnext, d)$
  \State $drpair \gets drpair - (F2 - F1)$
  \EndIf
  \EndFor

  \State
  \State $\arc{DR}[i] \gets drpair / (N * \frac{4}{3} \pi)$
  \EndFor
\end{algorithmic}
\hrule
\end{figure}

In this appendix, we present pseudocode to calculate $\arc{DR}(r)$ precisely in $O(N_{\rm g})$ time, applicable to survey areas with rectangular (Fig.~\ref{alg1}), cuboidal (Fig.~\ref{alg2}), circular (Fig.~\ref{alg3}), or spherical (Fig.~\ref{alg4}) shapes. The explanations of the algorithms are discussed in Section~\ref{sec:dr}. \red{A Python code based on these algorithms is published as doi:10.5281/zenodo.5201479, as developed on \url{https://github.com/chongchonghe/analytic-2pcf/tree/v1.0.0}.}

\end{document}